\documentclass[pdflatex,sn-mathphys-num]{sn-jnl}

\usepackage{graphicx}%
\usepackage{multirow}%
\usepackage{amsmath,amssymb,amsfonts}%
\usepackage{amsthm}%
\usepackage{mathrsfs}%
\usepackage[title]{appendix}%
\usepackage{xcolor}%
\usepackage{textcomp}%
\usepackage{manyfoot}%
\usepackage{booktabs}%
\usepackage{algorithm}%
\usepackage{algorithmicx}%
\usepackage{algpseudocode}%
\usepackage{listings}%

\usepackage{tcolorbox}

\definecolor{viola}{HTML}{4D2FB2}

\theoremstyle{thmstyleone}%
\theoremstyle{thmstyletwo}%

\theoremstyle{thmstylethree}%

\begin{document}

\title[Wikipedia and Grokipedia: A Comparison of Human and Generative Encyclopedias]{Wikipedia and Grokipedia: A Comparison of Human and Generative Encyclopedias}


\author[1]{\fnm{Ortal} \sur{Hadad}}

\author[2]{\fnm{Edoardo} \sur{Loru}}
\equalcont{These authors contributed equally to this work.}

\author[1]{\fnm{Jacopo} \sur{Nudo}}
\equalcont{These authors contributed equally to this work.}

\author[3]{\fnm{Anita} \sur{Bonetti}}

\author[1]{\fnm{Matteo} \sur{Cinelli}}

\author*[1]{\fnm{Walter} \sur{Quattrociocchi}}\email{walter.quattrociocchi@uniroma1.it}

\affil[1]{\orgdiv{Department of Computer Science}, \orgname{Sapienza University of Rome}, \orgaddress{\street{Viale Regina Elena 295}, \city{Rome}, \postcode{00161}, \country{Italy}}}

\affil[2]{\orgdiv{Department of Computer, Control and Management Engineering}, \orgname{Sapienza University of Rome}, \orgaddress{\street{Via Ariosto 25}, \city{Rome}, \postcode{00185}, \country{Italy}}}

\affil[3]{\orgdiv{Department of Communication and Social Research}, \orgname{Sapienza University of Rome}, \orgaddress{\street{Via Salaria 113}, \postcode{00198}, \city{Rome}, \country{Italy}}}


\abstract{We present a comparative analysis of Wikipedia and Grokipedia to examine how generative mediation alters content selection, textual rewriting, narrative structure, and evaluative framing in encyclopedic content.
We model page inclusion in Grokipedia as a function of Wikipedia page popularity, density of reference, and recent editorial activity. Inclusion is non-uniform: pages with higher visibility and greater editorial conflict in Wikipedia are more likely to appear in Grokipedia. For included pages, we distinguish between verbatim reproduction and generative rewriting. Rewriting is more frequent for pages with higher reference density and recent controversy, while highly popular pages are more often reproduced without modification.
We compare editing activity across the two platforms and estimate page complexity using a fitness–complexity framework to assess whether generative mediation alters patterns of editorial participation.
To assess narrative organization, we construct actor–relation networks from article texts using abstract meaning representation. Across multiple topical domains, including U.S. politics, geopolitics, and conspiracy-related narratives, narrative structure remains largely consistent between the two sources. Analysis of lead sections shows broadly correlated framing, with localized shifts in laudatory and conflict-oriented language for some topics in Grokipedia.
Overall, generative systems preserve the main structural organization of encyclopedic content, while affecting how content is selected, rewritten, and framed.}




\maketitle

\section*{Introduction}\label{sec1}
Wikipedia has long served as a reference system for encyclopedic content, where material is produced and revised through explicit practices of sourcing, discussion, and editorial control. Disagreement is managed through visible mechanisms such as citations, edit histories, and governance rules, which allow readers to track how content changes over time \cite{wilkinson2007cooperation,shi2019wisdom,yang2022map}. As a result, encyclopedic content on Wikipedia reflects an ongoing revision process, in which uncertainty and disagreement remain explicitly represented.
Large language model–based systems are increasingly used across a wide range of applications, including search, content summarization, question answering, and automated text production. More recently, they have also been applied to the generation and mediation of encyclopedic content. Systems such as Grokipedia differ from collaborative reference platforms in that they provide users with a single synthesized textual output. Processes such as source selection, disagreement, and revision are not exposed explicitly, but are resolved during text generation.
Prior work has shown that fluent LLM-generated text can increase user confidence in its factual reliability, even when explicit verification cues are absent \cite{troshani2021we,steyvers2025large,colombatto2025influence}. Compared to collaborative encyclopedic platforms, generative systems therefore convey signals related to sourcing, contestation, and revision in a different way.
Existing research has extensively examined Wikipedia as a collaborative knowledge system, with attention to editorial conflict \cite{yasseri2012dynamics,borra2015societal}, polarization \cite{yang2024polarization}, and governance structures \cite{beschastnikh2008wikipedian,burke2008mopping,leskovec2010governance}. By contrast, less is known about how generative systems mediate encyclopedic content at the level of selection, rewriting, narrative structure, and framing. In this study, we address this gap through a systematic comparison of Wikipedia and Grokipedia.
A large body of work on large language models, indeed, has focused on issues such as hallucinations, bias, alignment, and factual accuracy \cite{jo2023promise,ye2024benchmarking,ji2024aligner,kalai2025language}. 
These two strands of research followed largely separate trajectories. 
As a result, limited attention has been paid to how generative systems interact with established reference knowledge infrastructures and how this interaction affects their structural properties.

Recent work suggests that delegating evaluative and interpretive tasks to large language models has implications that extend beyond isolated errors or hallucinations. Even when model outputs align with human judgments at the surface level, the processes through which such judgments are produced can differ substantially \cite{quattrociocchi2025epistemological}. In particular, large language models can generate fluent and authoritative-looking assessments that resemble expert evaluations, while relying primarily on statistical associations rather than on explicit evidential or normative reasoning.
This divergence has been described as epistemia, referring to situations in which linguistic plausibility substitutes for epistemic evaluation \cite{loru2025simulation,quattrociocchi2025epistemological}. Empirical evidence for this phenomenon has been reported both in controlled evaluative tasks, where LLMs reproduce expert judgments while following different heuristics, and in simulation settings, where generative agents amplify salient features and introduce structural distortions relative to human behavior \cite{nudo2025generative,hu2025generative,gao2025take}.

While prior studies show that generative systems can alter how judgment and representation are produced, they leave open a related question: how does encyclopedic content change when generative models are used to assemble and present existing reference material, rather than to evaluate or simulate it? In particular, little is known about how content is selected, rewritten, and stabilized when generative systems mediate information drawn from human-curated sources.
In this work, we address this gap by examining Grokipedia as a generative mediation layer built on top of Wikipedia. Rather than focusing on factual correctness, we analyze how generative mediation affects content selection, textual rewriting, narrative structure, and evaluative framing.
This focus allows us to study generative systems at the level of encyclopedic organization and presentation, and to assess how mediation by generative models reshapes reference content beyond individual outputs.

Recent studies have examined the Grokipedia project through quantitative analyses of its content. This work shows that Grokipedia largely reproduces existing Wikipedia material through direct copying, rewriting, or paraphrasing, rather than generating new content \cite{yasseri2025similar,triedman2025did}. Grokipedia entries are typically longer and more verbose than their Wikipedia counterparts \cite{yasseri2025similar}, consistent with the tendency of large language models to prioritize fluency. Rewriting activity is more frequent for controversial or sensitive topics, where issues of neutrality and source reliability are prominent \cite{triedman2025did,mehdizadeh2025epistemic}. In addition, prior studies report differences in source usage, with generative content relying more heavily on non-academic or lower-reliability sources \cite{yasseri2025similar,triedman2025did,mehdizadeh2025epistemic,eibl2026grokipedia}.
While these studies provide important insights into textual similarity, verbosity, source selection, and framing, they focus primarily on content-level properties. Less attention has been paid to how generative mediation affects the structural organization of encyclopedic content and the processes through which it is selected, organized, and presented.

In this study, we address this gap through a systematic structural comparison of Wikipedia and Grokipedia across U.S. politics, geopolitics, and conspiracy-related topics. We analyze patterns of page inclusion, rewriting, editor participation, narrative organization, and evaluative framing. We find that content selection in Grokipedia is non-uniform, with greater emphasis on controversial pages with higher editorial activity. For included pages, rewriting is more extensive for more prominent topics and is less tightly coupled to explicit reference practices. At the same time, narrative structure, captured through actor–relation networks, remains largely similar across the two sources. The main differences emerge in evaluative language, where generative content exhibits more conflict-oriented and moralized framing.
Taken together, these results show that generative encyclopedic systems preserve the core narrative organization of existing reference content, while modifying how content is selected, rewritten, and framed.

\section*{Results}\label{sec2}

To examine how encyclopedic knowledge changes under generative mediation, we conduct a systematic, page-level comparison between Wikipedia and Grokipedia across three domains: U.S. politics, geopolitics, and conspiracy-related content. We analyze paired pages along four complementary dimensions: content selection, textual rewriting, narrative structure, and evaluative framing. This structured approach allows us to distinguish surface-level similarities from deeper changes, identifying where generative systems preserve properties of reference knowledge and where they modify them.
We first examine which Wikipedia pages are included and rewritten in Grokipedia, treating both as measurable outcomes of the generative process. Next, we reconstruct narrative networks to assess whether generative mediation alters the structural organization of actors and their relations. Finally, we examine content framing by measuring systematic differences in evaluative language across laudatory and conflict-related dimensions. Taken together, these analyses show a consistent pattern: generative mediation largely preserves the narrative structure of encyclopedic knowledge, while introducing systematic differences in content selection and in evaluative framing on salient pages.

\subsection*{Selection and Rewriting as Signals}\label{sec:selection}

We treat inclusion as the first point at which generative mediation operates: Grokipedia does not sample Wikipedia uniformly, but selects a subset of pages that are presented as encyclopedic content. As a second step, we focus on which of the included pages are subsequently rewritten by Grok relative to their original Wikipedia versions.

To characterize these two selection processes, we quantify them using logistic regression and examine how page-level features associated with popularity, number of references, and editorial activity relate both to the probability that a Wikipedia page appears in Grokipedia and to the probability that an included page is selected for generative rewriting by Grok. Wikipedia's English edition currently contains approximately 6 million articles and represents the largest global encyclopedia. By contrast, shortly after its launch in October 2025, Grokipedia included short of 900,000 English-language pages. Because nearly all Grokipedia entries can be traced back to Wikipedia pages with the same title, we identify not only which Wikipedia pages have a corresponding Grokipedia entry, but also which of these pages have been rewritten or kept unchanged. We consider a Grokipedia page to be not rewritten if it contains the standard Creative Commons footer. Specifically, we determine whether a Grokipedia article is kept unchanged by checking for the presence of the following text at the bottom of the page: ``The content is adapted from Wikipedia, licensed under Creative Commons Attribution-ShareAlike 4.0 License''.

For each Wikipedia page, we collect a set of page-level features capturing visibility, documentation, and editorial activity: monthly page views ($V$), used as a proxy for popularity; the number of references ($R$), reflecting the depth of documentation; the number of edits in the previous two months ($E$), capturing recent contributor activity; and the number of reverts in the same period ($Rv$), indicating editorial disagreement, as edits made by one contributor are undone by another \cite{yasseri2012dynamics}.

Given the heavy-tailed distributions of these variables, all features were discretized (see Methods) into four ordinal activity levels: \emph{Low}, \emph{Mid}, \emph{High}, and \emph{Very High}. This discretization limits the influence of extreme values and improves in the models both coefficient stability and interpretability.

\begin{figure}[h]
    \centering
    \includegraphics[width=1\textwidth]{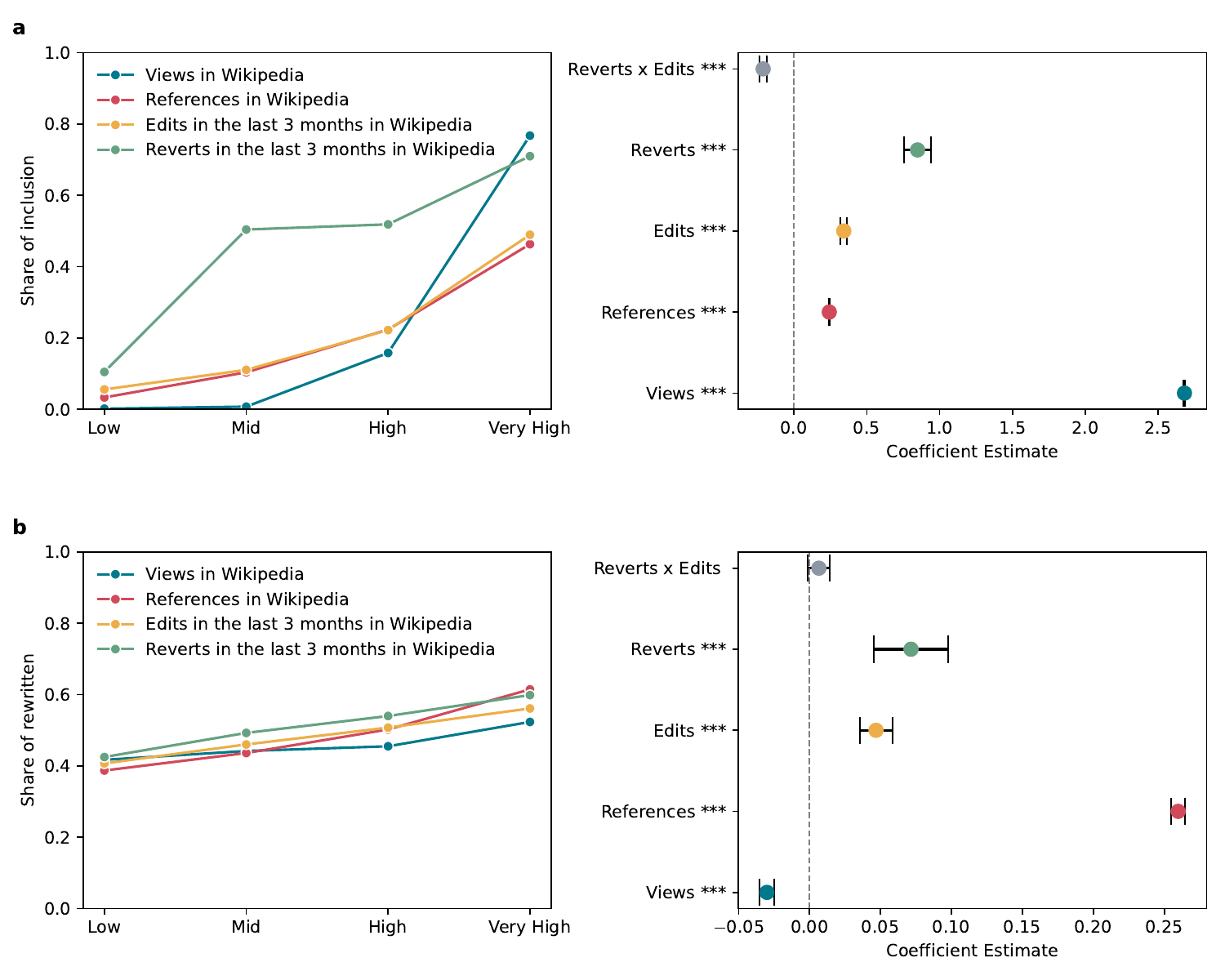}
    \caption{
    \textbf{Relationship between Wikipedia page characteristics and content selection and transformation on Grokipedia.}
    Predictors are discretized into four groups (\emph{Low}, \emph{Mid}, \emph{High}, and \emph{Very High}) and include page popularity (number of page views), content sourcing (number of references), and editorial activity (number of edits and number of reverts).
    Each row includes two plots: the first plot reports the share of articles exhibiting the outcome across predictor levels, while the second plot displays coefficient estimates from logistic regression models (coef-plots).
    The first row (\textbf{a}) shows the association between these characteristics and the probability that a Wikipedia page is included in Grokipedia.
    The second row (\textbf{b}) focuses on the probability that the Grokipedia version has been rewritten with respect to the original Wikipedia page.
    }
    \label{fig:inclusion}
\end{figure}

After discretization, we compute, for each activity level, the fraction of Wikipedia pages that are included in Grokipedia and, among the included pages, the fraction that are subsequently rewritten by Grok. The left panel of Fig. \ref{fig:inclusion}(a) reports inclusion probabilities, showing that the likelihood of a page appearing in Grokipedia increases monotonically with page popularity, reference density, and editorial activity. Pages with more attention, references, or editorial activity are likelier to be included.

The left panel of Fig. \ref{fig:inclusion}(b) presents an analogous analysis for generative rewriting, examining how the share of included pages that are rewritten by Grok varies across the same activity levels. The results show that the same features associated with inclusion are also systematically related to the probability of a page being selected for rewriting.

While these descriptive patterns are informative, a multivariate model is required to assess the relative contribution of each factor and to quantify their effects while controlling for confounding correlations. Let $Y_i$ denote a binary indicator equal to 1 if a Wikipedia page is present in Grokipedia and 0 otherwise, and let $Z_i$ denote a binary indicator equal to 1 if the corresponding Grokipedia page has been rewritten with respect to the original Wikipedia text and 0 if it reproduces it verbatim.
We decided to include the interaction term between \textit{reverts} ($Rv$) and \textit{edits} ($E$) because reverts are a special type of edits, and this interaction allows us to account for their correlation and better separate the effects of general edits from those specific to reverts.

Using the same set of predictors for both outcomes ($V$, $R$, $E$, $Rv$), and including the interaction term $E \cdot Rv$, we specify the following logistic regression model:
\begin{equation}
\log \frac{\Pr(Y_i \text{ or } Z_i = 1)}{1 - \Pr(Y_i \text{ or } Z_i = 1)} 
= \alpha_0 + \alpha_1 V_i + \alpha_2 E_i + \alpha_3 R_i + \alpha_4 Rv_i + \alpha_5 (E_i Rv_i)
\end{equation}
For inclusion ($Y_i$), coefficients represent the log-odds that a Wikipedia page is included in Grokipedia. Results are reported in Figure \ref{fig:inclusion}(a). Page popularity emerges as the strongest predictor: holding other variables constant, a page with \emph{Low} views has an estimated inclusion probability of 0.005, whereas pages with \emph{Very High} views reach an inclusion probability of 0.83. Editorial conflict, measured through the number of reverts, is also positively associated with inclusion, while reference density has a comparatively smaller effect. All estimated parameters are reported in Supplementary Table S1.

Having established the determinants of inclusion, we next examine the drivers of rewriting, focusing on which included pages Grokipedia intervenes in and modifies relative to their original Wikipedia versions.
Taking the rewritings \(Z_i\) into account: in Fig.~\ref{fig:inclusion}(b), the coefficients represent the \emph{log-odds} that a Grokipedia page differs from its corresponding Wikipedia entry. Holding other variables constant, page popularity is negatively associated with the probability of rewriting: highly viewed pages are more likely to be reproduced unchanged, while less visible pages are more often modified.
By contrast, the number of references emerges as the strongest predictor of rewriting. Pages with a higher number of references are significantly more likely to undergo AI-driven rewriting. Recent editorial activity, used as a proxy for ongoing attention and dispute, is also positively associated with rewriting, consistent with prior findings \cite{triedman2025did,eibl2026grokipedia}. Indeed, both edits and reverts have positive and statistically significant effects, indicating that Grokipedia intervenes more frequently on pages that remain subject to active revision and disagreement. Reverts, in particular, capture instances of editorial disagreement and are associated with a higher likelihood of rewriting.
The interaction between edits and reverts is positive but not statistically significant at conventional levels.
Complete results are reported in Supplementary Table S2.
This suggests that rewriting is associated with the presence of editorial conflict rather than the intensity of the editing activity.
Grokipedia tends to include highly visible pages, and among these, those characterized by higher instability and frequent reverts are more likely to be rewritten by Grok.

\subsection*{Community Editing Patterns and Page Complexity}\label{sec:complexity}

Similar to Wikipedia, registered users on Grokipedia can propose edits to existing pages, but the underlying review mechanisms differ substantially. On Wikipedia, editing is fully decentralized: contributors can directly implement changes and revert edits made by others. On Grokipedia, by contrast, an LLM serves as a centralized reviewer that evaluates proposed edits and decides on their acceptance, while also generating explanatory feedback.
Given this difference in editorial governance, it is important to assess whether editing patterns differ between the two platforms.
To address this question, we analyze edits proposed during the initial rollout of the editing feature on Grokipedia, between late October and November 2025 (see Methods), and compare them with publicly available Wikimedia data on revisions made on the corresponding set of pages during the same period. This comparison allows us to examine whether Grokipedia's editor community concentrates its activity on specific types of pages and to assess similarities and differences with editing behavior observed on Wikipedia. In particular, we restrict the analysis to the subset of 845 pages that are common to both platforms and that received at least two edits on each.

\begin{figure}[t] 
    \centering
    \includegraphics[width=1\textwidth]{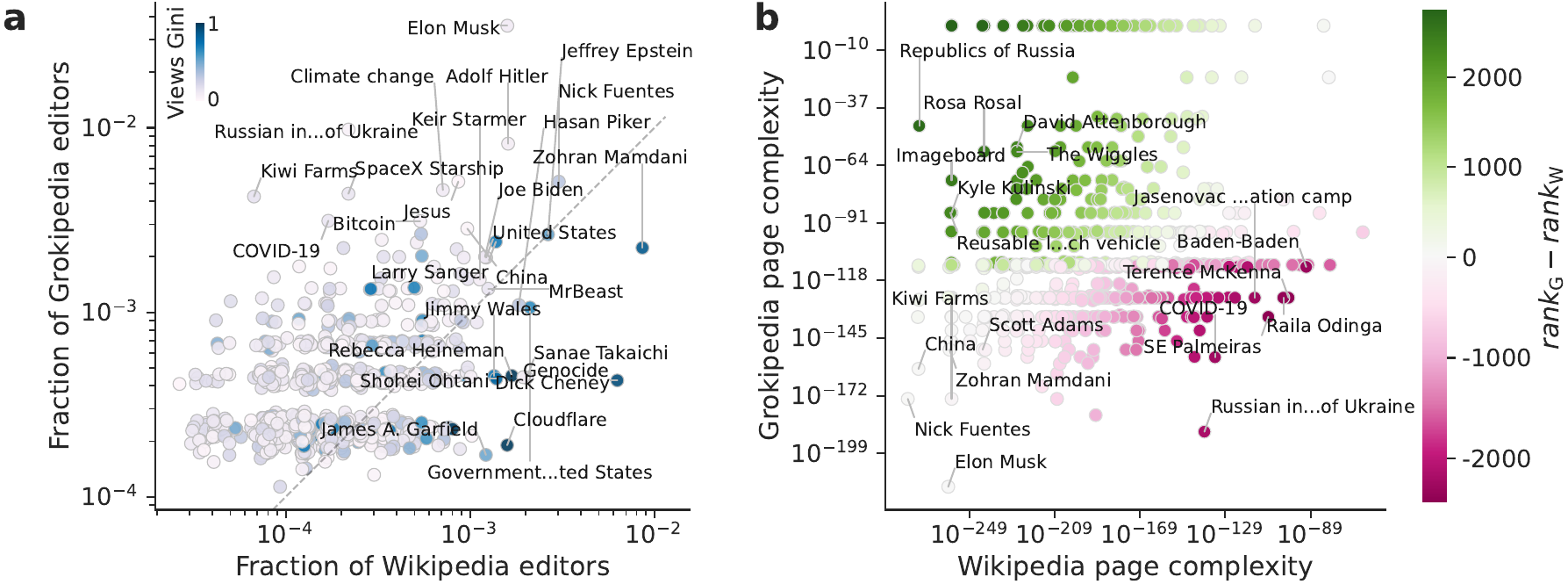} 
    \caption{\textbf{Editing dynamics on Grokipedia.} (a) Fractions of Grokipedia and Wikipedia editors who have contributed on a page. Only pages present in both datasets are shown, but the reported fraction is relative to each platform's entire set of pages. The dashed line corresponds to equal fractions. To reduce overplotting at low fractions, points are shown with a small amount of random jitter. Color indicates the Gini index of Wikipedia page views in November 2025, with higher values indicating greater concentration on specific days rather than uniformity across the month. (b) Page complexity on Grokipedia compared with Wikipedia, calculated from users' editing behavior. Color indicates the difference in ranking between the same page on Grokipedia and Wikipedia, ordered by increasing complexity and referring to the matching subset of pages only. Pages comparatively less (more) complex on Grokipedia are characterized by a negative (positive) difference in ranking.}
    \label{fig:edits}
\end{figure}

First, we measure, for each page, the fraction of editors on each platform who contributed at least one edit. Higher fractions indicate that a page attracts attention from a larger share of the contributor community. Examining pages at the extremes of this distribution, therefore, provides insight into the types of topics that motivate user participation on each platform. The results are reported in Fig. \ref{fig:edits}(a) and show clear differences between Wikipedia and Grokipedia.

On Wikipedia, pages with the highest shares of distinct contributors are primarily associated with topics that were highly salient during the observation window, often in connection with specific news events. We measure this concentration by computing the Gini index \cite{gini1921measurement} of Wikipedia page views in November 2025, where values closer to 1 indicate that views are highly concentrated on specific days. These events include the deaths of public figures (e.g., Dick Cheney and Rebecca Heineman), appointments to public office (e.g., Sanae Takaichi and Zohran Mamdani), and controversies directly involving Wikipedia itself, as reflected in the page related to Jimmy Wales. These pages have strong, event-driven concentrations of editing activity.

In contrast, the same pages attracted comparatively little attention from contributors on Grokipedia, who instead concentrated their activity on broader and more enduring topics that were not necessarily associated with short-term spikes in attention during the study period. Among the pages with the highest contributor fractions are entries related to Elon Musk and his companies, as well as historically or politically salient figures (e.g., Adolf Hitler and Vladimir Putin) and long-running controversial topics (e.g., COVID-19, climate change, the Russian invasion of Ukraine, and Kiwi Farms). Although Grokipedia was introduced only recently, and the focus on these pages may partly reflect an early-stage effect, these preliminary findings point to differences in how contributor attention is distributed across pages on the two platforms. 

We next extend this analysis by examining the composition of the editor communities associated with each page. In this context, pages differ not only in topic and visibility, but also in the type of contributors they attract. Some pages are edited by a broad and heterogeneous set of contributors, while others concentrate activity among a smaller and more specialized group. As a result, the composition of the editor community provides information about the level of expertise required to contribute to a page.
To capture this relationship between pages and their contributing editors, we adopt a measure of complexity introduced in prior work \cite{Tacchella2012}, originally developed to quantify the complexity of products based on the fitness of the countries that export them.

We adapt the fitness–complexity framework to Wikipedia and Grokipedia by interpreting editors as countries and pages as products, and by constructing an editor–page bipartite matrix indicating whether a given editor has contributed to a given page. Within this framework, fitness and complexity are defined by means of a recursive relationship according to which an editor's fitness reflects the breadth of their activity across pages of varying difficulty, while a page's complexity captures the level of expertise required to contribute to it.
The underlying intuition is straightforward. Highly active or broadly experienced editors tend to contribute across many pages, including both simple and complex ones; as a result, their presence alone provides limited information about a page's complexity. Conversely, when a page attracts contributions primarily from less active or less experienced editors, this suggests a lower level of complexity. Page complexity therefore emerges as a function of the fitness of its contributing editors and is reduced by the participation of low-fitness contributors. A detailed description of the method is provided in the Methods section.

In our setting, we apply the fitness–complexity framework to directly compare the complexity of the same pages on Grokipedia and Wikipedia. Differences in estimated page complexity thus indicate that the same topic is edited by contributor communities with different activity profiles across the two platforms. After computing page complexity separately for Grokipedia and Wikipedia, we restrict the analysis to the common subset of pages with at least two edits on each platform.
The results, shown in Fig. \ref{fig:edits}(b), indicate a weak, positive correlation between page complexity on Grokipedia and Wikipedia (Spearman's $\rho = 0.18$, $P < 0.001$). This suggests limited similarity in the composition of contributor communities editing the same pages across the two platforms. Focusing on individual pages, noteworthy outliers emerge. For instance, ``Russian invasion of Ukraine'' and ``Republics of Russia'' exhibit markedly different complexity profiles despite both addressing Russian geopolitics. The former has relatively low complexity on Grokipedia, and, as shown in Fig. \ref{fig:edits}(a), is also edited by a substantial fraction of Grokipedia editors. By contrast, the latter is considerably less complex on Wikipedia than on Grokipedia. Although the two pages differ substantially in complexity on both platforms, the reversal of their relative rankings suggests differences in the composition of the contributor communities across the two platforms. At the same time, pages such as ``Kiwi Farms'' rank among the least complex on both Grokipedia and Wikipedia, yet attract vastly different fractions of editors across the two platforms. This indicates that, despite differences in how many editors contribute, these pages draw on contributors with a similarly limited scope of activity, resulting in comparable levels of complexity on both platforms.

These findings are consistent with the presence of partially distinct editor communities supporting Grokipedia and Wikipedia. While contributors on the two platforms may exhibit similar aggregate editing behaviors, they appear to differ in how attention and effort are distributed across topics. 
We note, however, that the analysis of Grokipedia edits is based on a limited observation window shortly after the introduction of the editing feature. As a result, these patterns should be interpreted cautiously, and further analysis will be needed once a more extensive revision history becomes available.

\subsection*{Narrative Network Structure across Platforms}\label{sec:narrative-robustness}

Large language models generate text by recombining linguistic patterns learned from human-produced corpora, rather than by relying on explicit representations of the world. This raises a natural question: when encyclopedic content is rewritten by a generative system, are changes limited to surface form, or do they also affect how narratives are structured and presented? We address this question by comparing the narrative organization of Wikipedia articles with their generated counterparts on Grokipedia.

Narratives can play an important role in shaping content understanding by organizing events, actors, and actions into interpretable structures that convey meaning and evaluation \cite{velleman2003narrative, fulton2005narrative}. At the same time, measuring narratives computationally is challenging, as they are often implicit, heterogeneous, and fragmented, motivating diverse analytical approaches, including for the specific study of Wikipedia content \cite{shuttleworth2024pov, metilli2019wikidata}.

To address this challenge, we adopt a published operational approach that defines narratives in a minimal and tractable sense \cite{pournaki_willaert_2025}. Specifically, we define a narrative as a collection of relations between actors, specifying who acts upon whom and how that interaction is framed. At its core, each narrative element can be expressed as an actor–predicate–object triplet, capturing an action or relation between two entities. Importantly, such relations are not purely descriptive: they can convey support, opposition, or neutrality, thereby shaping evaluative framing.
To construct these narrative networks, we rely on Abstract Meaning Representation (AMR), which extracts normalized relational structures from text and abstracts away from surface-level linguistic variation. Using AMR, we identify core agent–target relations within sentences and assign evaluative polarity based on a curated mapping of relation types \cite{pournaki_willaert_2025}.

These actor–relation triplets can naturally be represented as a network. In this representation, entities correspond to nodes, and directed edges encode relations from agents to targets. Each edge can be assigned a polarity: neutral, supportive, or conflictive, reflecting the evaluative tone with which the relation is expressed. For example, the sentence ``In October 2025, Kelly credited Trump for facilitating an Israel–Hamas deal'', drawn from the Mark Kelly page in the U.S. politics corpus, is decomposed into the predicate ASCRIBE \cite{di_fabio_etal_2019_verbatlas} with arguments ARG0: Mark Kelly and ARG1: Donald Trump. This yields a directed, supportive edge from the node Mark Kelly to the node Donald Trump. Aggregating these relations across an article yields a structured representation of its narrative organization that is independent of specific wording or stylistic choices. We retain only entities associated with consistent semantic categories (such as persons, organizations, locations, or events), ensuring comparability across articles and platforms.
Our analysis focuses on a subset of highly visited Wikipedia pages, which capture sustained user interest rather than transient attention. We group these pages into three topical domains, U.S. politics, geopolitics, and conspiracy-related content, and restrict attention to pages that are shared between Wikipedia and Grokipedia. This yields a corpus of 185, 150, and 60 pages, respectively. For each domain, the resulting narrative representation takes the form of a directed, signed, and weighted multigraph, where edge direction encodes agent–target relations, sign captures evaluative framing, and edge weights reflect the frequency of relations across the corpus. For a breakdown of each network's number of nodes and edges, see Supplementary Table S3.

In both corpora, neutral relations constitute the majority of narrative links, indicating that descriptive associations are prevalent over explicitly evaluative framing (see Supplementary Table S4). All polarity-specific layers are sparse, with low directed edge densities, and degree distributions are highly unequal, reflecting a concentration of narrative connectivity among a limited set of actors (see Supplementary Table S4 and S5). Actor rankings show substantial consistency across sources, pointing to a shared backbone of prominent entities in each domain (see Supplementary Table S6). 

\begin{figure}[t] 
    \centering
    \includegraphics[width=1\textwidth, ]{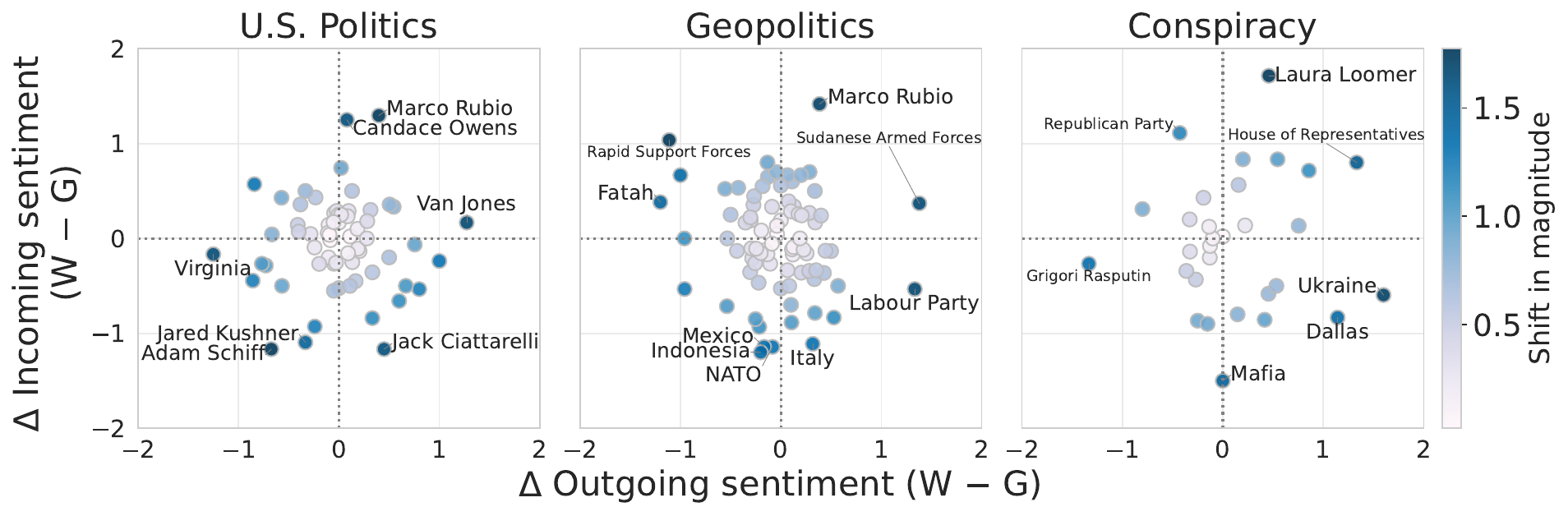} 
    \caption{\textbf{Actor-level differences in evaluative positioning between Wikipedia and Grokipedia across U.S. politics, geopolitics, and conspiracy narratives.} Each point corresponds to an actor with high evaluative activity, positioned by the difference in outgoing (x-axis) and incoming (y-axis) sentiment balance between Grokipedia and Wikipedia. Outgoing differences reflect changes in how actors evaluate others, whereas incoming differences reflect changes in how actors are evaluated by the surrounding narrative. Positive values indicate more supportive (or less conflictive) relations in Wikipedia relative to Grokipedia, while negative values indicate the opposite. Color encodes the magnitude of the overall displacement. For clarity, only actors with large displacements, corresponding to an absolute difference of at least 1 in either outgoing or incoming sentiment, are labeled.}
    \label{fig:narrative-shifts}
\end{figure}

Beyond local connectivity patterns, we examine higher-order structure within conflictive narratives, focusing on features that capture how opposition is organized at the network level. In this context, reciprocal relations correspond to mutual antagonism between pairs of actors, while cycles represent closed chains of conflict that may indicate sustained disputes or coalition-like antagonistic configurations. Across domains, we observe similarly low levels of both reciprocity and cyclic structure in Wikipedia and Grokipedia (see Supplementary Tables S7 and S8), suggesting that conflict is typically directed toward specific actors rather than organized into extended back-and-forth exchanges or factional patterns.
These results indicate that generative rewriting does not substantially alter narrative organization at the global level. Instead, Grokipedia largely preserves the actor–relation structure present in Wikipedia, providing a stable structural baseline against which more localized narrative differences can be examined.

To characterize actor-level evaluative positioning, we aggregate supportive and conflictive relations into directional sentiment balances, computed separately for outgoing and incoming interactions. Sentiment balance is defined as the difference between supportive and conflictive relation weights, normalized by their sum, yielding a signed measure that ranges from -1 to 1. Positive values indicate predominantly supportive relations, negative values predominantly conflictive relations, and values near zero reflect a balance between the two. Outgoing sentiment captures how an actor evaluates others, whereas incoming sentiment reflects how the actor is evaluated within the surrounding narrative.
Figure~\ref{fig:narrative-shifts} reports cross-source differences in actor-level sentiment balances for U.S. politics, geopolitics, and conspiracy-related narratives. Each point represents an actor, positioned by the difference between Wikipedia and Grokipedia in outgoing ($x$-axis) and incoming ($y$-axis) sentiment balance, with the origin corresponding to identical evaluative positioning across sources. Positive values indicate relatively more supportive (or less conflictive) relations in Wikipedia compared to Grokipedia, while negative values indicate the opposite. Across domains, more than 75\% of actors lie within a narrow region around the origin, indicating limited differences in evaluative positioning between the two sources. Larger displacements are confined to a small subset of actors, such as Marco Rubio and Candace Owens. These displacements typically occur along a single dimension (i.e., either outgoing or incoming sentiment), while simultaneous shifts along both dimensions are comparatively rare. Overall, these patterns are consistent with localized reframing in the generated corpus, in which evaluative positioning changes for specific actors without disrupting the overall narrative structure inherited from Wikipedia.

\begin{figure}[t] 
    \centering
    \includegraphics[width=1\textwidth]{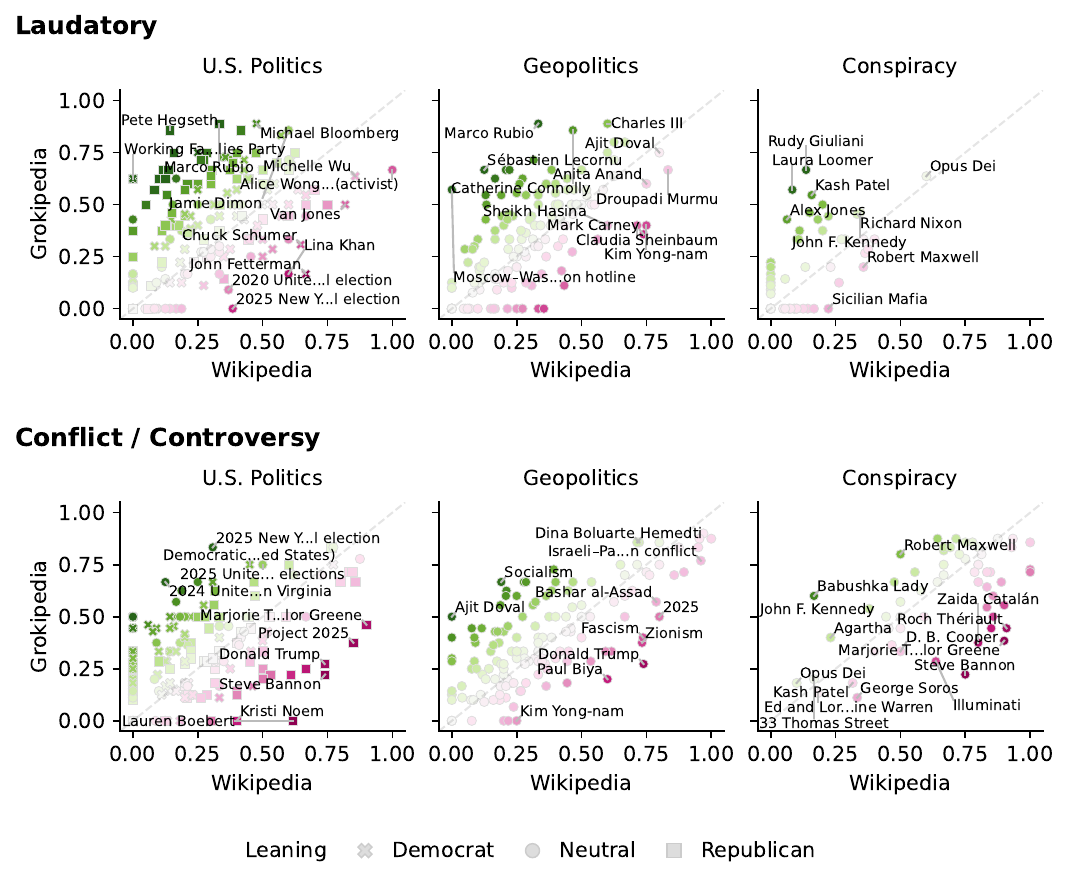} 
    \caption{\textbf{Content framing scores in Grokipedia and Wikipedia articles across U.S. Politics, Geopolitics, and Conspiracy-related pages.} Top: fraction of sentences in the lead section that show praise, admiration, or glorification. Bottom: fraction of sentences in the lead section that focus on disputes, disagreements, or controversies. Color intensity is proportional to the difference between the two fractions, while point shape for pages in U.S. Politics refers to their political leaning. The dashed line represents the quadrant bisector, corresponding to an equal fraction on both platforms. Only a subset of pages is labeled for visual clarity, and among these, some are shortened to improve readability. While scores tend to be weakly or moderately correlated, noteworthy outliers emerge, especially among U.S. Politics pages.}
    \label{fig:framing}
\end{figure}

\subsection*{Evaluative Framing under Generative Mediation}\label{sec:framing-drift}
Next, we analyze how U.S. Politics, Geopolitics, and Conspiracy-related pages present content from a framing perspective, focusing on two dimensions: (i) the extent to which a page adopts an explicitly laudatory framing, and (ii) the extent to which it emphasizes conflict or controversy beyond merely acknowledging their presence.
Rather than analyzing the entire article body, we restrict our attention to the so-called ``lead section'', which in Wikipedia corresponds to the introductory part of the article that precedes the first heading \cite{WikipediaManualStyleLead}. Although Grokipedia does not explicitly adopt this terminology, we apply the same criterion to identify the corresponding section in Grokipedia articles. This choice is motivated by prior research on reader behavior, which shows that the lead section is not only the first part encountered by visitors, but often also the only one they read \cite{wikimediaResearchWhichParts, Lamprecht2016, Lamprecht2021}. For this reason, even shifts in framing within this section alone can disproportionately affect readers' overall interpretation.

Content framing is assessed using an LLM-based annotation procedure. To obtain a quantitative score for each of the two framing dimensions, we first split each lead section into sentences and restrict the analysis to lead sections containing at least 500 characters. This filter excludes 21 pages from the initial set. For the remaining pages (170 for U.S. Politics, 145 for Geopolitics, and 59 for Conspiracy), we compute the fraction of sentences that the model classifies as exhibiting the corresponding framing. This yields a score in the interval $[0,1]$, where values close to $0$ indicate predominantly neutral framing and values close to $1$ indicate stronger alignment with the given dimension. The exact prompt and further details on the annotation procedure are reported in Methods.

The results shown in Fig.~\ref{fig:framing} indicate that, overall, framing patterns in Grokipedia are correlated with those observed in the corresponding Wikipedia entries. For laudatory framing, we observe Spearman correlations of $\rho = 0.47$ for U.S. Politics, $\rho = 0.65$ for Geopolitics, and $\rho = 0.54$ for Conspiracy-related pages. For Conflict / Controversy framing, the corresponding correlations are $\rho = 0.41$ for U.S. Politics, $\rho = 0.65$ for Geopolitics, and $\rho = 0.47$ for Conspiracy-related content. All correlations are statistically significant ($P < 0.001$), indicating substantial overall similarity in framing across the two platforms.
At the same time, Fig.~\ref{fig:framing} highlights a set of notable outliers, particularly within the U.S. Politics domain. For these pages, Grokipedia exhibits systematically more laudatory framing and a reduced emphasis on conflict or controversy relative to Wikipedia. Examples include political figures such as Donald Trump, Marjorie Taylor Greene, Marco Rubio, and Pete Hegseth, as well as affiliated actors and entities including Steven Bannon and Project 2025. A similar pattern is observed within the Conspiracy-related subset, where Grokipedia pages on Rudy Giuliani, Laura Loomer, and Alex Jones contain a higher fraction of sentences expressing laudatory framing.

\section*{Conclusions}\label{sec12}

This study compares Wikipedia and Grokipedia to examine how encyclopedic content changes under generative mediation, focusing on content selection, textual rewriting, narrative structure, and evaluative framing. The analysis reveals that Grokipedia does not sample Wikipedia uniformly. Pages with higher visibility and recent editorial activity are more likely to be included, while pages characterized by a high number of references and editorial conflicts are more frequently rewritten. In contrast, the overall narrative structure of articles, as captured through actor–relation networks, remains largely consistent across both platforms. The main differences are observed in evaluative framing, where Grokipedia exhibits localized shifts in laudatory and conflict-oriented language for a subset of salient topics.

The results indicate that generative encyclopedic systems largely preserve the structural organization of reference knowledge while intervening in content selection, rewriting, and consolidation. Differences between human and generative encyclopedias therefore pertain primarily to which pages are included and rewritten, and how contested content is stabilized into a single textual output, rather than to systematic changes in presented narratives or framing.
Our analysis, however, is limited to the structural and textual properties of encyclopedic content. Factual accuracy, persuasive effects, and ideological intent are not evaluated. The contribution of this work is a systematic empirical comparison of reference and generative encyclopedic systems, highlighting how generative mediation operates at the levels of selection, rewriting, and framing, while preserving much of the underlying narrative structure.
From a methodological perspective, these findings suggest that evaluations of generative knowledge systems based solely on accuracy or ideological bias are insufficient to capture their full impact. When generative systems serve as interfaces to reference knowledge, additional dimensions become relevant, including content selection from the reference corpus, consolidation of contested material through rewriting, and alterations in evaluative framing at the point of presentation. More broadly, these findings indicate that generative encyclopedic systems should be studied not only as the output of text generators, but also as mediating layers that reorganize existing knowledge infrastructures, a perspective that requires combining content-level comparison with structural analysis.

\section*{Methods}
\label{sec:methods}

\subsection*{Data collection}
\label{sec:data}

\subsubsection*{Grokipedia}
\label{sec:data_grokipedia}
We collected a snapshot of Grokipedia v0.1 between October 29 and November 02, 2025. We first retrieved the full list of available pages from Grokipedia's sitemap \cite{grokipedia_sitemap}, which at the time of collection included a total of 876,250 unique page URLs. Then, we downloaded the full HTML of each listed page using the \texttt{requests} library in Python \cite{githubrequests} via GET requests. This process yielded a final dataset of 874,955 downloaded pages, of which approximately 5\% contained no content beyond the message \textit{``This page doesn't exist... yet''}. The latter were thus excluded from our analyses.

Additionally, we gathered the full history of edits available up until November 24, 2025. These edits were collected by querying Grokipedia's \texttt{list-edit-requests-by-slug} API endpoint. Downloaded data includes both metadata about the edit, such as its author or creation time, and the suggested change, along with Grok's feedback. At the time of downloading, only a subset of pages featured at least one edit. The final dataset thus includes 13,116 proposed edits by 4,506 authors across 4,730 pages, between October 27 and November 24. We note that the dataset presents a window where only 104 edits were proposed, between November 4 and November 13 (see Supplementary Fig. S1). Two possible reasons can explain this drop: (i) after its initial rollout, the feature was temporarily restricted to a subset of contributors; (ii) users could propose edits during this time period, but they were only registered by the system at a later time.

\subsubsection*{Wikipedia}
\label{sec:data_wikipedia}
For each English Wikipedia page, we counted the number of page views collected during the first month of 2025 (January) as a measure of the page's popularity over a relatively recent period. These data were obtained from the publicly available page view dumps provided by the Wikimedia Foundation \cite{wikimedia_pageview}.

Then, for each page, we collected information about the number of references. Reference counts were obtained by downloading and parsing the full English Wikipedia content dump released on 20 November 2025 \cite{wikimediaEnwikiDump}, allowing us to consistently extract citation information across all pages.

Revision histories made to English Wikipedia pages were collected from the official dumps provided monthly by the Wikimedia Foundation \cite{wikimedia_mediawiki_history}. Specifically, we downloaded the dump released in November 2025, focusing only on revisions submitted between October and November 2025 (named \texttt{2025-11.enwiki.2025-10.tsv.bz2} and \texttt{2025-11.enwiki.2025-11.tsv.bz2}, respectively). Although the Wikimedia dump contains the complete revision history of each page, we restrict our analysis to the same time period as Grokipedia (October 27 to November 24) to enable a direct comparison. Editing activity can vary across time as contributors concentrate on topics that are particularly salient at a given moment. Aligning the Grokipedia and Wikipedia datasets temporally, therefore, allows us to compare editing behavior while controlling for topical salience.

For the narrative and framing analyses, we identify pages by collecting the 5{,}000 most-viewed English Wikipedia pages over the preceding 60 days (snapshot: 29 November 2025), computed by aggregating daily page view counts obtained via the Wikipedia Pageviews API. The resulting set defines the initial set of candidate pages. Article text, lead sections, and associated metadata are subsequently retrieved using the Wikipedia API.

\subsection*{Discretization of Wikipedia Activity Features}
\label{sec:discretization}
We considered several features extracted from Wikipedia pages, such as the number of page views, edits, reverts, and references. Since all these variables exhibit strongly right-skewed (heavy-tailed) distributions, using them in their raw continuous form would be problematic.  
To address this issue, we transformed each feature into four ordered activity levels: \emph{Low}, \emph{Mid}, \emph{High}, and \emph{Very High}. The discretization is performed using an iterative mean-based procedure \cite{glanzel1988characteristic}, which has been widely adopted in previous studies \cite{abramo2017investigation,cinelli2021ambiguity,di2024patterns}. The core idea is to progressively separate low and moderate values from increasingly extreme ones by repeatedly comparing observations to the mean of the remaining data.

Let $X$ be the set of values of a given continuous feature. At each iteration $k$, we compute the mean
\[
\mu^{(k)} = \frac{1}{|X^{(k)}|} \sum_{x \in X^{(k)}} x,
\]
where $X^{(k)}$ denotes the subset of values not yet assigned to a category.  
All values smaller than this mean are assigned to the current activity level and removed from the set. The procedure is repeated on the remaining values.

Formally, the four categories are defined as
\[
\begin{aligned}
\text{Low} &= \{ x \in X^{(0)} \mid x < \mu^{(0)} \}, \\
\text{Mid} &= \{ x \in X^{(1)} \mid x < \mu^{(1)} \}, \\
\text{High} &= \{ x \in X^{(2)} \mid x < \mu^{(2)} \}, \\
\text{Very High} &= X^{(3)},
\end{aligned}
\]
with the remaining set updated at each step as
\[
X^{(k+1)} = X^{(k)} \setminus \text{category}_k.
\]%
This procedure ensures that each successive category captures increasingly extreme values, reflecting the heavy-tailed nature of the original distribution.

\subsection*{Page complexity}
\label{sec:methods_page_complexity}
To estimate Grokipedia's and Wikipedia's page complexity, we apply a technique initially introduced by \cite{Tacchella2012} to assess countries' fitness (i.e., competitiveness) and products' complexity. The software implementation is publicly available as a Python module \cite{githubGitHubEFCdatafermi}. We use the same procedure as in the original formulation, but in our analysis, editors take the role of countries and pages take the role of products. In this representation, an editor's Fitness reflects the breadth and difficulty of the pages to which they contribute, while a page's Complexity reflects the level of expertise required to edit it. Because highly active or skilled editors tend to contribute across a wide range of pages, their participation alone carries limited information about page Complexity. In contrast, the presence of less fit editors is an indication of a page's low Complexity. This interpretation mirrors the original framework and makes the editor–page network well-suited to the Fitness–Complexity approach. Below is an outline of the employed algorithm. For its formal definition and a discussion of its properties, we refer the reader to \cite{Tacchella2012}.

The algorithm operates on a binary editor–page matrix $M$ where $M_{ep}=1$ indicates that editor $e$ has proposed at least one edit on page $p$. Then, the algorithm computes two coupled vectors: editor Fitness and page Complexity. All values are initially set to 1, and both vectors are updated iteratively until convergence.
At each iteration $n$, an editor's Fitness is computed as the sum of the Complexity values (from iteration $n-1$) of the pages they edited. Conversely, a page's Complexity is computed as the inverse of the sum of the inverse Fitness values (from iteration $n-1$) of the editors who contributed to it. At the end of the iteration, both vectors are normalized. After convergence, the final results are thus the Fitness value for each editor and the Complexity value for each page, with the latter serving as our metric of interest in this study.

\subsection*{Page domain categorization}
Starting from the set of 5{,}000 most-viewed pages on Wikipedia (see the Data collection section), we first filter for pages that exist on both Wikipedia and Grokipedia. Page titles are then classified by topic using a LLM–based classifier (Gemini 2.5 Flash), which operates solely on page titles. This initial classification assigns pages to several broad topical categories, with geopolitics, U.S. politics, and conspiracy-related topics as the target categories, alongside additional domains such as entertainment, sports, science, and history. We focus exclusively on current U.S. politics and geopolitics, as specified in the model instructions, defined as actors, institutions, and events active between 2015 and the present.

To obtain a more precise and conservative topic selection, we supplement this initial filtering with two additional instruction-tuned language models: \texttt{gemma-2-9b-it} and \texttt{llama-3.1-8b-Instruct}. For these models, classification is performed using both the page title and the Wikipedia lead section, truncated to a maximum of 1,000 characters. For each model, we apply two binary classifiers, identifying pages relevant to U.S. politics and to geopolitics, respectively. The prompts used for these classifications are provided in the Supplementary Information (see Supplementary Figures S2, S3, S4, and S5). Across the three topical domains, models identified a total of 440 unique pages as relevant to U.S. politics, 374 to geopolitics, and 382 to conspiracy-related content. Following manual validation and filtering, the final dataset consisted of 185 U.S. politics pages, 150 geopolitics pages, and 60 conspiracy-related pages. This human adjudication step ensures that borderline cases are consistently assigned.

\subsection*{Narrative networks}

We construct narrative networks for Wikipedia and Grokipedia using a network-based narrative framework adapted from \cite{pournaki_willaert_2025}. Narratives are operationalized through Abstract Meaning Representation (AMR) graphs, which enable the systematic extraction of entities and relations from text. All analyses are conducted separately for the three topical domains identified, namely current U.S. politics, current geopolitics, and conspiracy, to ensure domain-specific processing across platforms.

\subsubsection*{Network Construction}
  
For each domain and source, the full texts of the selected articles are processed through a sequential pipeline. Articles are first segmented into sentences using the Python module spaCy \cite{honnibal_montani_2020}, after which each sentence is parsed into an AMR graph using a pretrained neural parser from the IBM \texttt{transition-amr-parser} framework \cite{ibm_transition_amr_parser}. Parsed AMR graphs are thus decoded and transformed into predicate–argument representations following \cite{pournaki_willaert_2025}. From these representations, we extract predicate–ARG0–ARG1 tuples, retaining only well-formed instances encoding a clear agent–patient relation. Non-informative elements, including pronouns, numerals, and generic placeholders, are removed. 

Since we observe in the data that some of the same entities are referred to by different names, we apply a multi-stage entity normalization procedure. First, for each person appearing in the initial page set, we construct a mapping from family names to canonical full names based on corresponding Wikipedia page titles. Ambiguous family names shared by multiple entities (e.g., Trump) are disambiguated using page-level context, mapping mentions to the entity associated with the page on which they occur. In a second step, we manually curate high-frequency entities, defined as entities appearing in more than five sentences in the corpus, to correct residual abbreviations and aliases (e.g., mapping IDF to Israel Defense Forces). 

Entity semantic types are obtained directly from the AMR parser output: for each entity occurrence in ARG0 and ARG1 positions, the parser assigns a coarse semantic category (e.g., person, organization, country). For each platform, these assignments are aggregated across all occurrences of an entity, and the most frequently observed type is selected. We retain only entities that appear in both Wikipedia and Grokipedia with a consistent, non-empty semantic type. Entities for which the most frequent type differs across platforms, or for which no type is assigned, are excluded from further analysis.

Next, we aggregate the extracted relations into a narrative network in which nodes represent entities and edges capture narrative relations from ARG0 to ARG1. Each relation is assigned a polarity, either supportive, conflictive, or neutral, following the manually curated VerbAtlas-based\cite{di_fabio_etal_2019_verbatlas} mapping of \cite{pournaki_willaert_2025}. The frequency of each relation thus determines the edge weight, resulting in a directed, weighted, signed narrative multigraph for each platform and domain. 
To ensure cross-platform comparability, all subsequent analyses are restricted to a shared semantic backbone consisting of entities that appear in both Wikipedia and Grokipedia.

\subsubsection*{Sentiment Polarity}
   
Sentiment-based analyses are restricted to nodes with a sufficient number of signed relations (i.e., supportive or conflictive). For each node, we compute its sentiment mass as the total weight of supportive and conflictive edges, calculated separately for incoming and outgoing directions. Nodes are then ranked by sentiment mass within each corpus, and only nodes in the top decile are retained for subsequent analyses. For each retained node, we compute the normalized incoming and outgoing sentiment
balances, defined as the difference between supportive and conflictive edge weights divided by their sum (Eq.~\ref{eq:polarity}):

\begin{equation}
\begin{alignedat}{2}
\Delta S^{\mathrm{out}} &=
\frac{E^{\mathrm{out}}_{\mathrm{sup}} - E^{\mathrm{out}}_{\mathrm{con}}}
     {E^{\mathrm{out}}_{\mathrm{sup}} + E^{\mathrm{out}}_{\mathrm{con}}}
\qquad
&
\Delta S^{\mathrm{in}} &=
\frac{E^{\mathrm{in}}_{\mathrm{sup}} - E^{\mathrm{in}}_{\mathrm{con}}}
     {E^{\mathrm{in}}_{\mathrm{sup}} + E^{\mathrm{in}}_{\mathrm{con}}}
\end{alignedat}
\label{eq:polarity}
\end{equation}

\vspace{0.5\baselineskip}
Here, $E^{\mathrm{out}}_{\mathrm{sup}}$ and $E^{\mathrm{out}}_{\mathrm{con}}$
denote the total weights of outgoing supportive and conflictive edges of a given node, respectively, with analogous definitions for incoming edges. These balances define a two-dimensional role representation based on outgoing and incoming sentiment, which we compute independently for Wikipedia and Grokipedia and compare on the set of mutual nodes. Cross-platform narrative shifts, displayed in Fig. \ref{fig:narrative-shifts}, are thus quantified as displacements in this role space:

\begin{equation}
\label{eq:role_displacement}
\Big(
\Delta S^{\mathrm{out}}_{\text{W}} - \Delta S^{\mathrm{out}}_{\text{G}},
\;
\Delta S^{\mathrm{in}}_{\text{W}} - \Delta S^{\mathrm{in}}_{\text{G}}
\Big) 
\end{equation}

\subsection*{Content framing}
\begin{figure}[t]
\centering
\scalebox{1}{
\begin{tcolorbox}[
colback=gray!8,
colframe=white,
boxrule=0.5pt,
arc=5pt,
left=8pt,
right=8pt,
top=6pt,
bottom=6pt,
width=\textwidth
]
    You are an automated text analysis system.
    
    \vspace{1em}
    
    Task:
    
    Analyze the framing of the provided text snippet. Do NOT evaluate factual accuracy, topic, or intent beyond framing.
    
    \vspace{1em}
    
    Dimensions to evaluate:
    \begin{itemize}
        \item \texttt{laudatory\_framing}: Does the text praise, admire, or glorify something or someone?
        \item \texttt{conflict\_controversy}: Does the text focus on disputes, disagreements, or controversies, rather than merely mentioning them?
    \end{itemize}
    
    \vspace{0.5em}
    
    Output:
    
    Return only a valid JSON assigning 1 if the snippet exhibits the dimension, or 0 otherwise.
    
    \vspace{1em}
    
    Text: \texttt{\{LEAD SECTION SENTENCE\}}
    \end{tcolorbox}
    }
    \caption{Prompt used to assess content framing in an article's lead section.}
    \label{fig:prompt_framing}
\end{figure}

We use an LLM to analyze how text, specifically, the lead sections of Grokipedia's and Wikipedia's pages, is framed along two dimensions. To generate numerical scores, each lead section is split into sentences using the Python module spaCy \cite{honnibal_montani_2020} and model \texttt{en\_core\_web\_sm}. This analysis is restricted to lead sections with at least 500 characters. After subdividing all lead sections, each sentence is independently labeled by an LLM with the prompt in Fig. \ref{fig:prompt_framing}. We tested multiple models and, upon visual inspection of the results, found \texttt{gemma3:4b} to be the one better adhering to the used prompt. However, the other models' scores are moderately to highly correlated with those from \texttt{gemma3:4b} (see Supplementary Figs. S6 and S7). For this reason, all reported results were obtained through this model.

Queries are made locally via the interface provided by \texttt{ollama} and its Python module \cite{ollama}. We use all default parameters except for setting the temperature to 0, which enhances output reproducibility. Output format is enforced using the Structured Output functionality, which ensures that the LLM produces an output in JSON format with two predefined keys, \texttt{laudatory\_framing} and \texttt{conflict\_controversy}, each taking a value of either $0$ or $1$. This approach allows straightforward computation of a final score for both framing dimensions, by averaging the corresponding values across all sentences of that lead section.

Additionally, we label all pages in the U.S. Politics subset with their political affiliation, as reported in Fig. \ref{fig:framing}. For this task, we use the model \texttt{gemini-2.5-flash} with temperature 0 and with Thinking mode and Grounding with Google Search enabled, via the Google AI Studio web interface. Pages are labeled as ``Democrat'', ``Republican'', ``Neutral'', or ``Unsure'', with the latter provided by the model in case the person or entity is unknown. Therefore, we manually reviewed these classifications (16 out of the 185 total pages). The exact prompt used is provided in Supplementary Fig. S8. 

\section*{Data availability}
Grokipedia data can be freely retrieved via GET requests of the HTML pages and by querying the official API for edits. Wikipedia data can be retrieved from the dumps provided by the Wikimedia Foundation or via the official API. Processed data resulting from this study will be made available upon reasonable request to the corresponding author. 

\section*{Code availability}
The software implementation of the fitness–complexity algorithm is available as a Python module installable via \texttt{pip} (\href{https://github.com/EFC-data/fermi}{https://github.com/EFC-data/fermi}). The Python scripts used for narrative analysis are made available on a GitHub repository by the method's original authors \cite{pournaki_willaert_2025} (\href{https://github.com/pournaki/soteu-narratives}{https://github.com/pournaki/soteu-narratives}). All other analyses were carried out using base Python and freely available modules.

\section*{Author contribution}
O.H., E.L., J.N. collected the data; O.H., E.L., J.N., W.Q. designed the experiments; O.H., E.L., J.N. conducted the experiments; All authors wrote the manuscript and provided critical feedback.

\backmatter

\bmhead{Acknowledgements}
This work was supported by the SERICS project (PE00000014) under the NRRP MUR program funded by the European Union – NextGenerationEU; the DECODE project funded by the Presidency of the Council of Ministers (Italy); and the MUSMA project (PRIN 2022, CUP G53D23002930006) funded by the Italian Ministry of University and Research (MUR) under the EU NextGenerationEU initiative, Mission 4, Component 2, Investment 1.1.





\bibliography{sn-bibliography}

\end{document}